\documentclass[11pt,ams]{article}%
\usepackage{geometry}
\usepackage{amsmath}
\usepackage{amsfonts}
\usepackage{amssymb}
\usepackage{graphicx}%

\providecommand{\U}[1]{\protect\rule{.1in}{.1in}}

\begin{document}

\date{}
\title{\textbf{Equivalence between Zwanziger's horizon function and Gribov's no-pole ghost form factor }}
\author{\textbf{\ } \textbf{A.\thinspace J.\thinspace G\'{o}mez$^{a}$%
\thanks{ajgomez@uerj.br}}\thinspace, \textbf{M.~S.~Guimaraes}$^a$\thanks{marceloguima@gmail.com}, \textbf{R.\thinspace F.\thinspace
Sobreiro}$^{b}$\thanks{sobreiro@if.uff.br}\ , \textbf{S.\thinspace
P.\thinspace Sorella}$^{a}$\thanks{sorella@uerj.br}\ \thanks{Work supported by
FAPERJ, Funda{\c{c}}{\~{a}}o de Amparo {\`{a}} Pesquisa do Estado do Rio de
Janeiro, under the program \textit{Cientista do Nosso Estado},
E-26/100.615/2007.}\\\textit{$^{a}${\small {UERJ $-$ Universidade do Estado do Rio de Janeiro}}}\\\textit{{\small {Instituto de F\'{\i}sica $-$ Departamento de F\'{\i}sica
Te\'{o}rica}}}\\\textit{{\small {Rua S{\~a}o Francisco Xavier 524, 20550-013 Maracan{\~a}, Rio
de Janeiro, Brasil}}}\\\textit{\ }$^{b}$\textit{{\small {UFF $-$ Universidade Federal Fluminense,}}}\\
\textit{\small{Instituto de F\'{\i}sica, Campus da Praia Vermelha}}\\\textit{{\small {Avenida General Milton Tavares de Souza s/n, 24210-346, Niter\'oi, Brasil}}}}
\maketitle

\begin{abstract}
The ghost form factor entering the Gribov no-pole condition is evaluated till the third order in the gauge fields. The resulting expression turns out to coincide with Zwanziger's horizon function implementing the restriction to the Gribov region in the functional integral.
\end{abstract}


\section{Introduction}
In his seminal work \cite{Gribov:1977wm}, Gribov pointed out that the Landau gauge condition $\partial_\mu A^{a}_\mu=0$ is plagued by the existence of gauge copies, {\it i.e.} there exist equivalent configurations $A'_\mu = U A_\mu U^{-1} + i U \partial_{\mu} U^{-1} $ which still obey the condition, $\partial_{\mu} A'^{a}_{\mu}=0$.  As a consequence, the Landau gauge does not enable us to pick up a unique field representative for each gauge orbit\footnote{It is worth to point out that the existence of the Gribov copies is not restricted to the Landau gauge, being in fact a general feature of the gauge fixing procedure \cite{Singer:1978dk}.} . \\\\In order to get rid of the gauge copies, Gribov proposed \cite{Gribov:1977wm} to restrict the domain of integration in the Feynman path integral to a certain region $\Omega$, defined as the set of field configurations obeying the Landau condition and for which the Faddeev-Popov operator ${\cal M}^{ab}$, ${\cal M}^{ab}=-(\partial^2 \delta^{ab} -g f^{abc}A^{c}_{\mu}\partial_{\mu})$, is strictly positive, namely
\begin{equation}
\Omega \;= \; \{ A^a_{\mu}\;; \;\; \partial_\mu A^a_{\mu}=0\;; \;\; {\cal M}^{ab}=-(\partial^2 \delta^{ab} -g f^{abc}A^{c}_{\mu}\partial_{\mu})\; >0 \; \} \;. \label{gr}
\end{equation}
The boundary $\partial\Omega$ of the region $\Omega$,  where the first vanishing eigenvalue of the operator ${\cal M}^{ab}$ appears, is called the first Gribov horizon. One has to note that, within the region $\Omega$, the operator
${\cal M}^{ab}$ is strictly positive, so that its inverse $({\cal M}^{-1})^{ab}$ does exist. \\\\To restrict the domain of integration in the functional integral, Gribov worked out the so called no-pole condition \cite{Gribov:1977wm} for the ghost propagator, which is the inverse of the operator ${\cal M}^{ab}$, namely 
\begin{equation}
{\cal G}^{ab} = ({\cal M}^{-1})^{ab} \;, \label{new1}
\end{equation}
where the gauge field $A^{a}_\mu$ plays the role of an external classical field. Expression (\ref{new1}) can be represented in a functional form by means of the Faddeev-Popov ghosts 
\begin{equation}
{\cal G}^{ab}(x,y;A)   = \frac{1}{N^2-1} \langle \bar c^a(x) c^b(y) \rangle_{\rm conn} = \frac{1}{N^2-1} \frac{\int {\cal D} c{\cal D}\bar{c}\; \bar{c}^{a}(x)c^{b}(y)e^{-\int d^{4}x \bar{c}^{a}\partial_{\mu}D_{\mu}^{ab}c^{b}}}{\int {\cal D} c{\cal D}\bar{c}\; e^{-\int d^{4}x \bar{c}^{a}\partial_{\mu}D_{\mu}^{ab}c^{b}}} \;. \label{new2}
\end{equation}
According to \cite{Gribov:1977wm}, one introduces the ghost form factor $\sigma(k,A)$ in momentum space as 
\begin{equation}
{\cal G}(k;A) = \frac{1}{k^2} \frac{1}{1-\sigma(k,A)} \;, \label{new3}
\end{equation}
where ${\cal G}(k;A)$ is obtained by taking the Fourier transform of the trace of ${\cal G}^{ab}(x,y;A)$, {\it i.e.} 
\begin{equation}
{\cal G}(k;A) = \int d^4x d^4y \; e^{ik(x-y)} \; {\cal G}(x,y;A) \;, \label{new4}
\end{equation}
and 
\begin{equation}
{\cal G}(x,y;A) = Tr\; {\cal G}^{ab}(x,y;A) =  \sum_{a=1}^{N^2-1} {\cal G}^{aa}(x,y;A) \;. \label{new5}
\end{equation}
Before starting the evaluation of the form factor $\sigma(k,A)$ it is worth to point out that expression (\ref{new3}) can be obtained as the Fourier transform of the quantity 
\begin{equation}
{\cal G}^{*}(z;A) = \int d^4y \; {\cal G}(z,0;A_y) \;, \label{new6} 
\end{equation} 
where 
\begin{equation}
A_y(x) = A(x+y)  \;, \label{new7}
\end{equation}
{\it i.e.}
\begin{equation}
{\cal G}(k;A) = \int d^4z \; e^{ikz} \; {\cal G}^{*}(z;A) \;. \label{new7}
\end{equation}
This property\footnote{We are grateful to the referee for having pointed out eq.(\ref{new7}).} can be obtained from  
eq.(\ref{new4}) by performing the change of variables $(z=x-y, y=y)$, amounting to rewrite eq.(\ref{new4})  as 
\begin{equation}
{\cal G}(k;A) = \int d^4z \; e^{ikz} \int d^4y \; {\cal G}(z+y,y;A) \;. \label{new8}
\end{equation}
Finally, eq.(\ref{new7}) follows by observing that a translation of both arguments of ${\cal G}(z+y,y;A)$ by $y$ is the same as a translation of the field configuration $A^a_\mu$ by $y$, as it can be checked term by term by looking at the expressions given in the next sections, see for example eq.(\ref{zw6}). \\\\As $\sigma(k,A)$ turns out to be a decreasing function of the momentum $k$ \cite{Gribov:1977wm}, Gribov required the validity of the condition
\begin{equation}
\sigma(0,A) \le 1\;, \label{npc}
\end{equation}
which is known as the no-pole condition. From condition (\ref{npc}) it follows that the ghost propagator has no poles at finite values of the momentum $k$. Therefore, expression (\ref{new3}) stays always positive, meaning that the Gribov horizon $\partial \Omega$ is never crossed. The only allowed pole is at $k=0$, whose meaning is that of approaching the horizon $\partial \Omega$, where the ghost propagator is singular, due to the appearance of zero modes of the operator ${\cal M}^{ab}$. According to the no-pole prescription, eq.(\ref{npc}), the Faddeev-Popov quantization formula gets modified as \cite{Gribov:1977wm}
\begin{align}
d\mu_{FP}   & = {\cal D}A\; \delta(\partial A)\; det({\cal M}^{ab})\; e^{-S_{YM}}  \\ \nonumber
           & \Rightarrow  {\cal D}A\; \delta(\partial A)\; det({\cal M}^{ab})\; \theta(1-\sigma(0,A)) \; e^{-S_{YM}}  \;, \label{fpm}
\end{align}
where $S_{YM}$ is the Yang-Mills action
\begin{equation}
S_{YM} = \frac{1}{4} \int d^4x \; F^a_{\mu\nu}F^a_{\mu\nu} \;, \label{ym}
\end{equation}
and $\theta(x)$
stands for the step function. Making use of the integral representation
\begin{equation}
\theta(x) = \int_{-i\infty +\varepsilon}^{+i\infty +\varepsilon} \frac{d\beta}{2\pi i\beta} e^{-\beta x} \;, \label{theta}
\end{equation}
it turns out that the ghost form factor $\sigma(0,A)$ can be brought into the exponential of the Yang-Mills measure $d\mu_{FP}$, {\it i.e.}
\begin{equation}
e^{-S_{YM}} \Rightarrow \; e^{-(S_{YM}+\beta\sigma(0,A))}  \;. \label{ymm}
\end{equation}
We see thus that the Yang-Mills action gets modified by the addition of the factor $\sigma(0,A)$. Therefore, for the
partition function $\cal Z$, one writes
\begin{equation}
{\cal Z} = \int {\cal D}A\; \frac{d\beta}{2\pi i\beta}\; \delta(\partial A)\; det({\cal M}^{ab})\;  e^{-S_{YM}} \;e^{\beta(1-\sigma(0,A))} \;. \label{pf}
\end{equation}
Further, the integration over $\beta$ was evaluated by a saddle point approximation \cite{Gribov:1977wm}, yielding
\begin{equation}
{\cal Z} = {\cal N}\int {\cal D}A\; \delta(\partial A)\; det({\cal M}^{ab})\;  e^{-(S_{YM}+\beta^{*}\sigma(0,A))} \;, \label{pfsp}
\end{equation}
with $\beta^{*}$ determined by the gap equation \cite{Gribov:1977wm}
\begin{equation}
1= \frac{3Ng^2}{4} \int \frac{d^4k}{(2\pi)^4} \frac{1}{k^4+\frac{g^2N}{2(N^2-1)}{\beta^{*}}} \;. \label{gge}
\end{equation}
Independently, Zwanziger \cite{Zwanziger:1989mf,Zwanziger:1992qr} implemented the restriction to the Gribov region $\Omega$ by following a different route, based on the study of the smallest eigenvalue, $\lambda_{min}(A)$, of the Faddeev-Popov operator. Relying on the equivalence between the canonical and microcanonical ensembles in the infinite volume limit, he was able to show that the restriction to the Gribov region can be achieved by adding to the Yang-Mills action a nonlocal term $S_{h}$, known as the horizon function \cite{Zwanziger:1989mf,Zwanziger:1992qr}, namely
\begin{align}
S_{h}  & = \int d^4x\; h(x) \; \nonumber \\
 & =  g^{2}\int d^{4}x\;d^{4}y\; f^{abc}A_{\mu}^{b}(x)\left[ {\cal M}^{-1}\right]  _{xy}
^{ad}f^{dec}A_{\mu}^{e}(y)   \;. \label{11}
\end{align}
The resulting partition function cut-off at the Gribov horizon turns out to be
\begin{equation}
\int_{\Omega} {\cal D}A\; \delta(\partial A)\; det({\cal M}^{ab})\;  e^{-S_{YM}} \; = \;
\int {\cal D}A\; \delta(\partial A)\; det({\cal M}^{ab})\;  e^{-(S_{YM}+\gamma^4 S_{h})} \;, \label{zw1}
\end{equation}
where the massive parameter $\gamma$ is a dynamical parameter determined in a self-consistent way through the horizon condition \cite{Zwanziger:1989mf,Zwanziger:1992qr}
\begin{equation}
\left\langle h(x)   \right\rangle =4\left(  N^{2}-1\right) \;. \label{hc}
\end{equation}
To the first order, condition (\ref{hc}) reads
\begin{equation}
1= \frac{3Ng^2}{4} \int \frac{d^4k}{(2\pi)^4} \frac{1}{k^4+2g^2N\gamma^4} \;, \label{zge}
\end{equation}
from which one sees that, apart from a numerical coefficient,  $\gamma^4$ can be identified with $\beta^{*}$, {\it i.e.} $\beta^*=4(N^2-1)\gamma^4$.  \\\\Although both Gribov's no pole condition  (\ref{npc}) and Zwanziger's construction of the horizon function $S_{h}$ amount to modify the Faddeev-Popov functional measure,  a discussion about the equivalence between the ghost form factor $\sigma(0,A)$ and the horizon function $S_{h}$ has not yet been worked out. The present work aims at filling this gap. We shall evaluate the form factor $\sigma(0,A)$ till the third order in the gauge fields $A^a_{\mu}$. The resulting expression will be thus compared with that obtained by expanding the horizon function $S_{h}$, hence establishing the equivalence between $\sigma(0,A)$ and $S_{h}$ till the third order in the gauge field expansion. \\\\The paper is organized as follows. In Sect.2 we evaluate Gribov's ghost form factor $\sigma(0,A)$. In Sect.3 we expand the horizon function $S_{h}$ by comparing it with $\sigma(0,A)$. Sect.4 is devoted to a few concluding remarks.

\section{Evaluation of Gribov's ghost form factor}

\label{Eq}

The evaluation of the ghost form factor $\sigma(k,A)$ will be performed order by order in the gauge field $A^{a}_\mu$. As we shall evaluate $\sigma$ to the third order, we write
\begin{align}
\sigma = \sigma^{(1)} + \sigma^{(2)} + \sigma^{(3)} + O(A^4) \;, \label{gg2}
\end{align}
where $\sigma^{(1)},   \sigma^{(2)}, \sigma^{(3)}$ stand, respectively,  for the first, second and third order expansion of $\sigma$ in powers of the gauge fields. Therefore, from the no-pole condition  (\ref{npc}), we get
\begin{equation}
{\cal G}(k;A) = \frac{1}{k^2} \left( 1+ \sigma^{(1)} + \sigma^{(2)} + \sigma^{(3)} + 2\sigma^{(1)}\sigma^{(2)}+ \sigma^{(1)} \sigma^{(1)} +  \sigma^{(1)} \sigma^{(1)} \sigma^{(1)} +O(A^4) \right) \;. \label{gg3}
\end{equation}
Let us start thus by considering the expression of  ${\cal G}(x,y;A)$ in an external background gauge field  $A^a_{\mu}$, obtained by taking the trace over the color indices of expression (\ref{new2}), namely 
\begin{equation}
{\cal G}(x,y;A)   = \frac{1}{N^2-1} \langle \bar c^a(x) c^a(y) \rangle_{\rm conn} = \frac{1}{N^2-1} \frac{\int {\cal D} c{\cal D}\bar{c}\; \bar{c}^{a}(x)c^{a}(y)e^{-\int d^{4}x \bar{c}^{a}\partial_{\mu}D_{\mu}^{ab}c^{b}}}{\int {\cal D} c{\cal D}\bar{c}\; e^{-\int d^{4}x \bar{c}^{a}\partial_{\mu}D_{\mu}^{ab}c^{b}}} \;. \label{G}
\end{equation}
In order to evaluate ${\cal G}(x,y;A)$ till the third order in the gauge field $A^a_{\mu}$, we consider
\begin{align}
\int {\cal D}c{\cal D}\bar{c}\; \bar{c}^{a}(x)c^{a}(y)\; & e^{-\int d^{4}x\bar{c}%
^{a}\partial_{\mu}D_{\mu}^{ab}c^{b}}
= \int {\cal D} c{\cal D}\bar{c}\; \bar{c}^{a}(x)c^{a}(y)\left(  1+g\int d^{4}z_{1}%
\partial_{\mu}^{z_{1}}\bar{c}^{a_1}(z_{1})f^{a_{1}b_{1}c_{1}}A_{\mu}^{b_{1}%
}(z_{1})c^{c_{1}}(z_{1})\right. \nonumber\\
&  +\frac{1}{2}g^{2}\int d^{4}z_{1}d^{4}z_{2}\partial_{\mu}^{z_{1}}\bar{c}%
^{a_1}(z_{1})f^{a_{1}b_{1}c_{1}}A_{\mu}^{b_{1}}(z_{1})c^{c_{1}}(z_{1}%
)\partial_{\nu}^{z_{2}}\bar{c}^{a_2}(z_{2})f^{a_{2}b_{2}c_{2}}A_{\nu}^{b_{2}%
}(z_{2})c^{c_{2}}(z_{2})\nonumber\\
&  +\frac{1}{6}g^{3}\int d^{4}z_{1}d^{4}z_{2}d^{4}z_{3}\partial_{\mu}^{z_{1}%
}\bar{c}^{a_1}(z_{1})f^{a_{1}b_{1}c_{1}}A_{\mu}^{b_{1}}(z_{1})c^{c_{1}}%
(z_{1})\partial_{\nu}^{z_{2}}\bar{c}^{a_2}(z_{2})f^{a_{2}b_{2}c_{2}}\nonumber\\
&  \left.  \times A_{\nu}^{b_{2}}(z_{2})c^{c_{2}}(z_{2})\partial_{\lambda
}^{z_{3}}\bar{c}^{a_3}(z_{3})f^{a_{3}b_{3}c_{3}}A_{\lambda}^{b_{3}}%
(z_{3})c^{c_{3}}(z_{3})+......\right) \; e^{-\int d^4{x} {\bar c}^a \partial^2 c^a } \;. \label{q}%
\end{align}
To the zeroth order approximation, it turns out that
\begin{align}
{\cal G}^{(0)}(x,y;A)  = \frac{1}{(N^2-1)} \left\langle \bar{c}^{a}(x)c^{a}(y)\right\rangle
^{(0)}  = G_{0}(x-y) = \int\frac{d^{4}q}{(2\pi)^{4}}\frac{e^{iq(x-y)}}{q^{2}} \;. \label{14}%
\end{align}

\subsection{First order}

At first order we have
\begin{align}
{\cal G}^{(1)}(x,y;A) = \frac{1}{N^2-1} \left\langle \bar{c}^{a}(x)c^{a}(y)\right\rangle ^{(1)} \;. \label{f1}
\end{align}
Using
\begin{equation}
\left\langle \bar{c}^{a}(x)c^{b}(y)\right\rangle ^{(0)}=\delta^{ab}G_{0}(x-y) \;,
\label{z}
\end{equation}
we obtain
\begin{align}
{\cal G}^{(1)}(x,y;A)=\int d^{4}z_{1}G_{0}(x-z_{1})\partial_{\mu}^{z_{1}}G_{0}%
(z_{1}-y)f^{aba}A_{\mu}^{b}(z_{1}) \;. \label{g1}
\end{align}
Moreover, due to
\begin{align}
f^{aba}=0 \;, \label{fz}
\end{align}
it follows that ${\cal G}^{(1)}$ vanishes identically
\begin{align}
{\cal G}^{(1)}(x,y;A)=0 \;,  \label{g1z}
\end{align}
so that
\begin{align}
\sigma^{(1)} =0 \;. \label{gg1z}
\end{align}

\subsection{Second order}

Performing Wick contractions and using eq.(\ref{z}), one obtains
\begin{align}
{\cal G}^{(2)}(x,y;A)=-\frac{g^2}{(N^2-1)} f^{a_{1}b_{1}c_{1}}f^{a_{1}b_{2}c_{1}}\int d^{4}z_{1}%
d^{4}z_{2}G_{0}(x-z_{1})\partial_{\mu}^{z_{1}}G_{0}(z_{1}-z_{2})\partial_{\nu
}^{z_{2}}G_{0}(z_{2}-y)A_{\mu}^{b_{1}}(z_{1})A_{\nu}^{b_{2}}(z_{2}) \;. \nonumber
\\ \label{gg2}
\end{align}
Taking the Fourier transformation  of the expression above
\begin{align}
{\cal G}(k;A)=\int d^{4}x\;d^{4}y\; e^{ik(x-y)} {\cal G}(x,y;A) \;, \label{ft}
\end{align}
it follows
\begin{equation}
{\cal G}^{(2)}(k;A)=-\frac{Ng^2}{(N^2-1)} \int d^{4}z_{1}d^{4}z_{2}d^{4}xd^{4}ye^{ik(x-y)}G_{0}%
(x-z_{1})\partial_{\mu}^{z_{1}}G_{0}(z_{1}-z_{2})\partial_{\nu}^{z_{2}}%
G_{0}(z_{2}-y)A_{\mu}^{a}(z_{1})A_{\nu}^{a}(z_{2}) \;, \label{w}%
\end{equation}
where we have used the property
\begin{align}
f^{abc}f^{ebc}=N\delta^{ae} \;. \label{ff}
\end{align}
Setting
\begin{equation}
A_{\mu}^{a}(x)=\int\frac{d^{4}q}{(2\pi)^{4}}e^{iqx}A_{\mu}^{a}(q) \;, \label{r}%
\end{equation}
we obtain
\begin{equation}
{\cal G}^{(2)}(k;A)= \frac{Ng^2}{(N^2-1)}  \frac{1}{k^{4}}\int\frac{d^{4}q}{(2\pi)^{4}%
}\frac{(-k_{\nu})q_{\mu}}{q^{2}}A_{\mu}^{a}(-q-k)A_{\nu}^{a}(q+k) \;, \label{d}%
\end{equation}
which can be rewritten as
\begin{equation}
{\cal G}^{(2)}(k;A)=\frac{Ng^2}{(N^2-1)} {k^{4}}\int\frac{d^{4}q}{(2\pi)^{4}}\frac{k_{\mu}k_{\nu}%
}{(k-q)^{2}}A_{\mu}^{a}(-q)A_{\nu}^{a}(q) \ .\label{f}%
\end{equation}
Therefore, till the second order, for the no-pole condition we get
\begin{align}
{\cal G}(k;A) =  {\cal G}^{(0)}(k;A) + {\cal G}^{(2)}(k;A)   =\frac{1}{k^{2}}\left(  1+\sigma^{(2)}(k,A)\right)  \;,
\label{gg}
\end{align}
where
\begin{align}
\sigma^{(2)}(k,A) & =\frac{Ng^2}{(N^2-1)} \frac{k_{\mu}k_{\nu}}{k^{2}} \; I_{\mu\nu}(k)  \;, \label{ik} \\
I_{\mu\nu}(k) & = \int \frac{d^4q}{(2\pi)^4} \frac{A^a_{\mu}(-q) A^a_{\nu}(q)}{(k-q)^2} \;. \label{ik1}
\end{align}
Owing to the transversality of the gauge field $A^a_{\mu}(q)$
\begin{align}
q_{\mu} A^a_{\mu}(-q) A^a_{\nu}(q) = q_{\nu} A^a_{\mu}(-q) A^a_{nu}(q) =0 \;, \label{tr}
\end{align}
we can set
\begin{align}
A^a_{\mu}(-q) A^a_{\nu}(q)  & = \omega(A) \left( \delta_{\mu\nu} - \frac{q_\mu q_\nu}{q^2} \right )\;, \nonumber \\
\omega(A) & = \frac{1}{3} A^a_{\lambda}(-q) A^a_{\lambda}(q) \;. \label{om}
\end{align}
Thus
\begin{align}
I_{\mu\nu}(k) = \frac{1}{3} \int \frac{d^4q}{(2\pi)^4} \frac{A^a_{\lambda}(-q) A^a_{\lambda}(q)}{(k-q)^2}  \left( \delta_{\mu\nu} - \frac{q_\mu q_\nu}{q^2} \right ) \;. \label{ik2}
\end{align}
Following \cite{Gribov:1977wm}\footnote{See also ref.\cite{Sobreiro:2005ec}.}, it turns out that
\begin{align}
I_{\mu\nu}(0)  = \frac{1}{3} \int \frac{d^4q}{(2\pi)^4} \frac{A^a_{\lambda}(-q) A^a_{\lambda}(q)}{q^2}  \left( \delta_{\mu\nu} - \frac{q_\mu q_\nu}{q^2} \right ) \;
 = \frac{1}{4} \delta_{\mu\nu} \int \frac{d^4q}{(2\pi)^4} \frac{A^a_{\lambda}(-q) A^a_{\lambda}(q)}{q^2}  \;, \label{1k3}
\end{align}
so that for the Gribov no-pole form factor $\sigma^{(2)}$, one obtains
\begin{align}
\sigma^{(2)}(0,A) = \frac{Ng^2}{4(N^2-1)}  \int \frac{d^4q}{(2\pi)^4} \frac{A^a_{\lambda}(-q) A^a_{\lambda}(q)}{q^2}
\;. \label{sigma2}
\end{align}
Expression ({\ref{sigma2})  corresponds to the original Gribov approximation \cite{Gribov:1977wm},
and is equivalent to set  $\frac{1}{\partial D}\approx\frac{1}{\partial^{2}}$  in the horizon
funtion $\left(  \ref{11}\right)  $.

\subsection{3${{}^\circ}$ order}
To the third order
\begin{align}
{\cal G}^{(3)}(x,y;A)  &  =\frac{1}{6}f^{a_{1}b_{1}c_{1}}f^{a_{2}b_{2}c_{2}}%
f^{a_{3}b_{3}c_{3}} \frac{g^{3}}{N^2-1} \int d^{4}z_{1}d^{4}z_{2}d^{4}z_{3}A_{\mu}^{b_{1}%
}(z_{1})A_{\nu}^{b_{2}}(z_{2})A_{\lambda}^{b_{3}}(z_{3})\\
&  \times\left\langle \bar{c}^{a}(x)c^{a}(y)\partial_{\mu}^{z_{1}}\bar{c}%
^{a_1}(z_{1})c^{c_{1}}(z_{1})\partial_{\nu}^{z_{2}}\bar{c}^{a_2}(z_{2})c^{c_{2}%
}(z_{2})\partial_{\lambda}^{z_{3}}\bar{c}^{a_3}(z_{3})c^{c_{3}}(z_{3}%
)\right\rangle \;. \label{3o1}
\end{align}
Performing all possible Wick contractions and proceeding as in the case of ${\cal G}^{(2)}$, one finds
\begin{align}
G^{(3)}(x,y;A)  &  =
{\cal F}
^{b_{1}b_{2}b_{3}} \frac{g^{3}}{N^2-1}\int d^{4}z_{1}d^{4}z_{2}d^{4}z_{3}A_{\mu}^{b_{1}%
}(z_{1})A_{\nu}^{b_{2}}(z_{2})A_{\lambda}^{b_{3}}(z_{3})G_{0}(x-z_{1})\\
&  \times\partial_{\mu}^{z_{1}}G_{0}(z_{1}-z_{2})\partial_{\nu}^{z_{2}}%
G_{0}(z_{2}-z_{3})\partial_{\lambda}^{z_{3}}G_{0}(z_{3}-y) \;, \label{3o2}
\end{align}
where we have defined
\begin{align}
{\cal{F}}^{b_{1}b_{2}b_{3}}\equiv f^{a_{1}b_{1}a}f^{a_{2}b_{2}a_{1}}f^{ab_{3}a_{2}} \;. \label{cf}
\end{align}
Taking the Fouries transformation
\begin{align}
{\cal G} ^{(3)}(k;A)  &  =\int d^{4}xd^{4}y\;e^{ik\cdot(x-y)} {\cal G}^{(3)}(x,y;A) \\ &  =
\frac{{\cal{F}}^{b_{1}b_{2}b_{3}}}{N^2-1}\frac{i^{3}g^{3}}{k^{4}}\int\frac{d^{4}q_{5}}{(2\pi)^{4}
}\frac{d^{4}q_{6}}{(2\pi)^{4}}A_{\mu}^{b_{1}}(-q_{5}-k)A_{\nu}^{b_{2}}
(q_{5}-q_{6})A_{\lambda}^{b_{3}}(q_{6}+k)\frac{\left(  q_{5}\right)  _{\mu
}\left(  q_{6}\right)  _{\nu}\left(  -k_{\lambda}\right)  }{q_{5}^{2}q_{6}^{2}} \;, \label{ft3}
\end{align}
and using the transversality condition  $q_{\mu}A_{\mu}^{a}(q)=0$, one gets
\begin{align}
{\cal G}^{(3)}(k;A)  &  =i^{3}g^{3} {\frac{{\cal F}^{b_{1}b_{2}b_{3}}}{N^2-1}}\frac{k_{\mu}k_{\lambda}}{k^{4}}\int\frac{d^{4}q_{1}}
{(2\pi)^{4}}\frac{d^{4}q_{2}}{(2\pi)^{4}}A_{\mu}^{b_{1}}(-q_{1})A_{\nu}%
^{b_{2}}(q_{1}-q_{2})A_{\lambda}^{b_{3}}(q_{2})\frac{\left(  q_{2}-k\right)
_{\nu}}{\left(  q_{1}-k\right)  ^{2}\left(  q_{2}-k\right)  ^{2}}\\
&  =i^{3}g^{3} \frac{{\cal{F}}^{b_{1}b_{2}b_{3}}}{N^2-1}\frac{k_{\mu}k_{\lambda}}{k^{4}}\; I^{b_1b_2b_3}_{\mu\lambda}(k) \;. \label{ft4}
\end{align}
Proceeding as in the previous case, for the Gribov ghost form factor till the third order  we find
\begin{align}
\sigma^{(3)}(0;A)  &  =\lim_{k\rightarrow0}\left[  i^{3}g^{3} \frac{{\cal F}^{b_{1}b_{2}b_{3}}}{N^2-1}\frac{k_{\mu}k_{\lambda}}{k^{2}}\;I^{b_1b_2b_3}_{\mu\lambda}(k)\right] \\
&  =i^{3}g^{3} \frac{{\cal{F}}^{b_{1}b_{2}b_{3}}}{N^2-1}\frac{1}{4}\int\frac{d^{4}q_{1}}{(2\pi)^{4}}\frac{d^{4}%
q_{2}}{(2\pi)^{4}}A_{\mu}^{b_{1}}(-q_{1})A_{\nu}^{b_{2}}(q_{1}-q_{2})A_{\mu
}^{b_{3}}(q_{2})\frac{\left(  q_{2}\right)  _{\nu}}{\left(  q_{1}\right)
^{2}\left(  q_{2}\right)  ^{2}} \;, \label{ft4}
\end{align}
where use has been made of
\begin{align}
I^{b_1b_2b_3}_{\mu\lambda}(0)=\delta_{\mu\lambda} \frac{1}{4}\int\frac{d^{4}q_{1}}{(2\pi)^{4}}\frac{d^{4}q_{2}}%
{(2\pi)^{4}}A_{\sigma}^{b_{1}}(-q_{1})A_{\nu}^{b_{2}}(q_{1}-q_{2})A_{\sigma
}^{b_{3}}(q_{2})\frac{\left(  q_{1}\right)  _{\nu}}{\left(  q_{1}\right)
^{2}\left(  q_{2}\right)  ^{2}} \;. \label{io}
\end{align}

\section{Expansion of the horizon function}
In order to make a comparison between Gribov's ghost form factor $\sigma(0,A)$ and Zwanziger's horizon function $S_{h}$, we need to expand the expression
\begin{align}
S_{h}=g^{2}\int d^{4}xd^{4}yf^{abc}A_{\mu}^{b}(x)\left[  {\cal M}^{-1}\right]
_{xy}^{ad}f^{dec}A_{\mu}^{e}(y) \;, \label{zw1}
\end{align}
till the third oder in the gauge field $A^{a}_{\mu}$. To that end we evaluate
the inverse of the Faddeev-Popov operator  ${\cal M}^{-1}$,  which is equivalent to solve the problem
\begin{equation}
\left(  -\partial^{2}\delta^{ab}+gf^{abc}A_{\mu}^{c}\partial_{\mu
}\right)  G^{bd}(x,y)=\delta^{ad}\delta^{(4)}(x-y) \;, \label{zw2}
\end{equation}
where the Green function $G^{ab}(x,y)$ is evaluated as a series in the coupling constant\footnote{Notice that an expansion in $g$ is equivalent to an expansion in the gauge field $A^a_\mu$.} $g$, $\it i.e.$
\begin{equation}
G^{bd}(x,y)=G_{0}^{bd}(x-y)+gG_{1}^{bd}(x,y)+g^{2}G_{2}^{bd}(x,y)+....  \;, \label{zw3}
\end{equation}
where
\begin{align}
-\partial^{2}\delta^{ab}G_{0}^{bd}(x-y)=\delta^{ad}\delta^{(4)}(x-y) \;. \label{zw4}
\end{align}
Thus, at first order, we get
\begin{align}
 \left(  -\partial^{2}\delta^{ab}+  gf^{abc}A_{\mu}^{c}\partial_{\mu
}\right) & \left(  G_{0}^{bd}(x-y)+  gG_{1}^{bd}(x,y)+ O(g^{2})\right)
=\delta^{ad}\delta^{(4)}(x-y) \;,  \nonumber \\
-g\partial^{2}\delta^{ab}G_{1}^{bd}(x-y) & +gf^{abc}A_{\mu}^{c}
\partial_{\mu}G_{0}^{bd}(x-y)    =0 \;, \nonumber \\
\partial^{2}G_{1}^{ad}(x-y)  &  =f^{abc}A_{\mu}^{c}\partial_{\mu}
G_{0}^{bd}(x-y) \;, \label{zw5}
\end{align}
which gives
\begin{align}
G_{1}^{ad}(x,y)=\int d^{4}z\frac{1}{\left\vert x-z\right\vert ^{2}}
f^{abc}A_{\mu}^{c}(z)\partial_{\mu}^{z}\frac{\delta^{bd}}{\left\vert
z-y\right\vert ^{2}} \;. \label{zw6}
\end{align}
Therefore
\begin{align}
\left[  {\cal M}^{ad}\right]  ^{-1}=\frac{\delta^{ad}}{\left\vert x-y\right\vert
^{2}}+gf^{adc}\int d^{4}z\frac{1}{\left\vert x-z\right\vert ^{2}}A_{\mu}
^{c}(z)\partial_{\mu}^{z}\frac{1}{\left\vert z-y\right\vert ^{2}} \;. \label{zw7}
\end{align}
Consequently, till the third order in the gauge fields $A^a_{\mu}$, for the horizon function we obtain
\begin{align}
S_{h}  &  =\int d^{4}xd^{4}yf^{abc}A_{\mu}^{b}(x)\left(  \frac{\delta^{ad}
}{\left\vert x-y\right\vert ^{2}}+gf^{adc}\int d^{4}z\frac{1}{\left\vert
x-z\right\vert ^{2}}A_{\mu}^{c}(z)\partial_{\mu}^{z}\frac{1}{\left\vert
z-y\right\vert ^{2}}\right)  f^{dec}A_{\mu}^{e}(y) \nonumber \\
&  =g^{2}f^{abc}f^{aec}\int d^{4}xd^{4}yA_{\mu}^{b}(x)\frac{1}{\left\vert
x-y\right\vert ^{2}}A_{\mu}^{e}(y) \nonumber \\ & +g^{3}f^{adm}f^{abc}f^{dec}\int d^{4}
xd^{4}yd^{4}zA_{\mu}^{b}(x)\frac{1}{\left\vert x-z\right\vert ^{2}}A_{\nu}
^{m}(z)\partial_{\nu}^{z}\frac{1}{\left\vert z-y\right\vert ^{2}}A_{\mu}
^{e}(y) + O(A^4)\;. \label{zw8}
\end{align}
Finally, moving to the Fourier space,
\begin{equation}
S_{h}=g^{2}N\int\frac{d^{4}q}{(2\pi)^{4}}A_{\mu}^{b}(-q)\frac{1}{q^{2}}A_{\mu
}^{b}(q)+ig^{3}f^{adm}f^{abc}f^{dec}\int\frac{d^{4}q_{1}}{(2\pi)^{4}}
\frac{d^{4}q_{2}}{(2\pi)^{4}}A_{\mu}^{b}(-q_{1})A_{\nu}^{m}(q_{1}-q_{2}
)A_{\mu}^{e}(q_{2})\frac{\left(  q_{2}\right)  _{\nu}}{q_{1}^{2}q_{2}^{2}} \;. \label{zw9}
\end{equation}
Recalling now the expression for the ghost form factor $\sigma(0,k)$ till the third order, namely
\begin{align}
\sigma(0,k)  &  =\sigma^{(2)}(0,A)+\sigma^{(3)}(0,A)   \nonumber \\
&  =\frac{g^2}{4(N^2-1)}\left( N\int\frac{d^{4}q}{(2\pi)^{4}} \frac{A_{\mu}^{b}
(-q)A_{\mu}^{b}(q)}{q^2}-ig
{\cal{F}}^{b_{1}b_{2}b_{3}}\int\frac{d^{4}q_{1}}{(2\pi)^{4}}\frac{d^{4}q_{1}}
{(2\pi)^{4}}A_{\mu}^{b_{1}}(-q_{1})A_{\nu}^{b_{2}}(q_{1}-q_{2})A_{\mu}^{b_{3}
}(q_{2})\frac{\left(  q_{2}\right)  _{\nu}}{q_{1}^{2}q_{2}^{2}}\right)   \nonumber \\
\label{zw10}
\end{align}
and making use of
\begin{align}
{\cal{F}} ^{b_{1}b_{2}b_{3}} A_{\mu}^{b_{1}}(-q_{1})A_{\nu}^{b_{2}}(q_{1}-q_{2})A_{\mu}^{b_{3}
}(q_{2}) = - f^{adm}f^{abc}f^{dec}A_{\mu}^{b}(-q_{1})A_{\nu}^{m}(q_{1}-q_{2})A_{\mu}^{e}(q_{2})
 \;,\label{zw11}
\end{align}
it is apparent that, apart from a global factor, the expression of the Gribov ghost factor $\sigma(0,A)$ coincides with that obtained by expanding the horizon function till the same order\footnote{It is useful to remark here the identity $\beta^{*} \sigma(0,A)= \gamma^4 S_{h} + O(A^4)$.} , {\it i.e.}
\begin{align}
\sigma(0,A) = \frac{1}{4(N^2-1)} S_{h}  + O(A^4) \;. \label{zw12}
\end{align}

\section{Conclusion}

\label{Con}

In this work the equivalence between Gribov's ghost form factor $\sigma(0,A)$ and Zwanziger's horizon function $S_{h}$ has been investigated. The form factor $\sigma(0,A)$ has been evaluated till the third order in the gauge fields $A^{a}_{\mu}$ and proven to be equivalent with the horizon function $S_{h}$, as expressed by eq.({\ref{zw12}). Our result can be interpreted as a strong indication of the fact that Zwanziger's horizon function $S_{h}$ is an all orders resummation of Gribov's form factor $\sigma(0,A)$. \\\\Let us conclude by mentioning that, although being non-local, the horizon function $S_h$ can be cast in local form by means of the introduction of a suitable set of auxilairy fields. Remarkably, the resulting action turns out to be renormalizable to all orders \cite{Zwanziger:1989mf,Zwanziger:1992qr,Maggiore:1993wq,Dudal:2005na,Gracey:2006dr,Dudal:2007cw,Dudal:2008sp}.
\bigskip

\section*{Acknowledgments}

We thank D. Dudal, M. Huber, N. Vandersickel and D. Zwanziger for useful comments. \\\\The Conselho Nacional de Desenvolvimento Cient\'{\i}fico e Tecnol\'{o}gico
(CNPq-Brazil), the Faperj, Funda{\c{c}}{\~{a}}o de Amparo {\`{a}} Pesquisa do
Estado do Rio de Janeiro, the Latin American Center for Physics (CLAF), the
SR2-UERJ, The Proppi-UFF and the Coordena{\c{c}}{\~{a}}o de Aperfei{\c{c}}oamento de Pessoal
de N{\'{\i}}vel Superior (CAPES) are gratefully acknowledged for financial support.


\begin{thebibliography}{9}                                                                                                %


\bibitem {Gribov:1977wm}V. N. Gribov, Nucl. Phys. B \textbf{139}, (1978) 1

\bibitem{Singer:1978dk}
  I.~M.~Singer,
  Commun.\ Math.\ Phys.\  {\bf 60} (1978) 7.


\bibitem{Zwanziger:1989mf}
  D.~Zwanziger,
  Nucl.\ Phys.\  B {\bf 323} (1989) 513.


\bibitem{Zwanziger:1992qr}
  D.~Zwanziger,
  Nucl.\ Phys.\  B {\bf 399}, 477 (1993).


\bibitem{Sobreiro:2005ec}
  R.~F.~Sobreiro and S.~P.~Sorella,
  arXiv:hep-th/0504095.


\bibitem{Maggiore:1993wq}
  N.~Maggiore and M.~Schaden,
  Phys.\ Rev.\  D {\bf 50}, 6616 (1994)
  [arXiv:hep-th/9310111].


\bibitem{Dudal:2005na}
  D.~Dudal, R.~F.~Sobreiro, S.~P.~Sorella and H.~Verschelde,
  Phys.\ Rev.\  D {\bf 72}, 014016 (2005)
  [arXiv:hep-th/0502183].

\bibitem{Gracey:2006dr}
  J.~A.~Gracey,
  JHEP {\bf 0605}, 052 (2006)
  [arXiv:hep-ph/0605077].

\bibitem{Dudal:2007cw}
  D.~Dudal, S.~P.~Sorella, N.~Vandersickel and H.~Verschelde,
  Phys.\ Rev.\  D {\bf 77}, 071501 (2008)
  [arXiv:0711.4496 [hep-th]].

\bibitem{Dudal:2008sp}
  D.~Dudal, J.~A.~Gracey, S.~P.~Sorella, N.~Vandersickel and H.~Verschelde,
  Phys.\ Rev.\  D {\bf 78}, 065047 (2008)
  [arXiv:0806.4348 [hep-th]].

\end{thebibliography}
\end{document}